\documentclass[smallextended]{svjour3}       

\smartqed  
\usepackage{amsmath}
\usepackage{amsfonts}
\usepackage{txfonts}
\usepackage{stmaryrd}
\usepackage{amssymb}
\usepackage{graphics,epsfig}
\begin{document}

\title{On the Quantum Structure}



\author{Dong-Sheng Wang}


\institute{D.-S. Wang \at
              School of Physics, Shandong
University, Jinan 250100, China \\
              \email{wdsn1987@gmail.com}}

\date{Received: date / Accepted: date}

\maketitle

\begin{abstract}

Quantum mechanics is a special kind of description of motion. The
concept of {\em wave function} itself implies the {\em openness} of
quantum system. We show that quantum mechanics describes the quantum
correlation, i.e., entanglement, and information in a new kind of
space, {\em tangnet} $\mathbb{T}^2$, where exist the basic {\em
quantum structure} of qubit and the universal {\em out-in} symmetry.
This work tries to form a new view to the fundamental problems of
the foundation of quantum mechanics.

\keywords {Quantum Structure \and Openness \and Tangnet \and Out-in
Symmetry}


\end{abstract}

\section{Introduction}
\label{intro}

In this work, we address the basic concepts of quantum mechanics
(QM) relating to the great development both on the foundation and
application these years. Besides the mathematical and experimental
aspects, the physical concepts of QM need particular attentions,
e.g., there are still lots of confusions of the ``weirdness'' of QM
at present. From the seminal work of EPR \cite{epr}, Schr\"odinger
\cite{Schrodinger}, Bohm \cite{bohm}, Bell \cite{Bell} etc, the
methods of {\em entanglement} and {\em nonlocality} have been widely
studied. Here, we do not focus on the confusions and differences
between the two methods; instead, we study the physical essence of
entanglement, quantum information, and further the new views of
``quantum''. Physically, the generalization from superposition to
entanglement is nontrivial. On one hand, it offers new ideas on what
superposition is; on the other hand, it leads to the
re-consideration of what quantum means. The existence of
entanglement has led to the growth of the fields of quantum
information and quantum computation (QIQC) \cite{Nielsen} and
quantum foundation (QF). In the research of QF, roughly speaking,
there are mainly two research trends: one is the interpretation of
quantum mechanics (IQM) \cite{schlosshauer}, such as the many-world
interpretation \cite{everett}; the other is the post-quantum
mechanics (PQM), such as the general probabilistic theory
\cite{Barrett}. In both IQM and PQM, entanglement plays the basic
role. In this short work, we do not intend to give the detailed
analysis of various theories; instead, we further investigate the
meaning of entanglement, information, and quantum from a new point
different with the present ones.

In Sec. \ref{sec:1}, we begin from the well known concept of
``openness'', and we discuss the strict physical meaning of wave
function. Then we introduce the method of ``quantum structure'' in
Sec. \ref{sec:2} which is the generalization of quantum state and
entanglement. We study the physical role of mutual information under
the spirit of openness. Also, we introduce the new space,
``tangnet'', where information is commonly shared, which is
particularly demonstrated by QM. Last, in Sec. \ref{sec:3} we
briefly analyze several related issues and open problems.

\section{Openness}
\label{sec:1}

In this section we discuss the concept of {\em openness} in the
study of quantum open system. Many quantum processes are due to the
openness of quantum system. For example, the lifetime of
micro-particle, the decay of electron from the excited state to
ground state etc, those are due to the interaction with the vacuum,
which cannot be removed. When the system is coupled with the
uncontrolled environment, decoherence will occur, which is described
by Zurek as the disturbance of system to environment \cite{Zurek}.
The role of openness in QM had been demonstrated a lot, such as the
early work of Zeh \cite{zeh}. Further, if we take a historical view,
one will find that early in the formation of the theory of density
matrix, the openness had already been addressed, and the concept of
{\em mixed state} was introduced \cite{springer}. We should note
that the approach of decoherence relies on density matrix. Below, we
analyze the meaning of mixed state. Generally, there are two related
views of mixed state, as follows\\

(I). The ``tracing'' view: {\em The mixed state is an inner part of
a global pure state, tracing out the rest.}\\

(II). The ``summing'' view: {\em The mixed state is the mixture of
several pure states, as $\rho=\sum_i p_i
|\psi_i\rangle\langle\psi_i|$, $|\psi_i\rangle$ is
pure state, $\sum_i p_i=1$.}\\

We should address that there is no standard reason why there should
be the two views, also, whether there could be anything more.
Mathematically, we can easily show that the two views are
equivalent. For the mixed state, according to the tracing view,
introducing the parameter $p_{ij}=\sum_{\mu}a_{i\mu}a_{j\mu}^*$,
then
\begin{eqnarray}
\label{eq:gf} \rho &=&\sum_{ij\mu}a_{i\mu}a_{j\mu}^*|i\rangle
\langle j|
\\ \nonumber
   &=&\sum_{ij\mu \nu}|\nu\rangle \langle \nu| a_{i\mu}a_{j\mu}^* |i\rangle \langle
   j| \\ \nonumber
   \end{eqnarray}
   \begin{eqnarray}
   & \nonumber \Rightarrow & \sum_{ij\mu \nu}|\mu\rangle \langle \nu| a_{i\mu}a_{j\nu}^* |i\rangle \langle
   j| \\ \nonumber
   &=& \sum_{i\mu}\sum_{j \nu} a_{i\mu}a_{j\nu}^* |i\rangle |\mu\rangle \langle j| \langle \nu|    \\ \nonumber
   &=& |\Psi \rangle \langle \Psi|,
\end{eqnarray}
where $|\Psi\rangle=\sum_{i\mu}a_{i\mu}|i\rangle|\mu\rangle$ is the
global pure state relating to $\rho$, the normalization rule
$\sum_{i\mu}|a_{i\mu}|^2=1$.

In practice, the differences between the two views are seldom
noticed, which can be explained by the mathematical equivalence.
However, for the physical meaning, the two views are different, and
we emphasize that the summing view (II) is wrong. From the tracing
view (I), we know that in reality we often cannot decide the wave
function of a system, as the system is often correlated with
environment (or other systems). Thus, as assumption, we use the
density matrix to describe the state of the system; that is, density
matrix is only the ``fragment'', which is not a complete
characterization of a state. In other words, there does not exist
several pure states, or ``relative state'' using Everett's
terminology, to mix up, that is, the summing view is in conflict
with the tracing view, physically. Further, taking from another
point, the summing view (II) of the mixed state actually comes from
the method of classical statistical physics. Classically, we can
say, e.g., the gas is the mixture of molecules. Yet, according to
QM, the concept of ``mixture'', which relates to the concept of
``classicality'', does not capture the feature of coherence, which
leads to, e.g., interference. So, it is not proper to introduce the
concept of mixture to QM, instead, we should study quantum process
from pure quantum methods, e.g., decoherence, and to avoid any
confusion with classicality.

\begin{figure}
\includegraphics[scale=0.4]{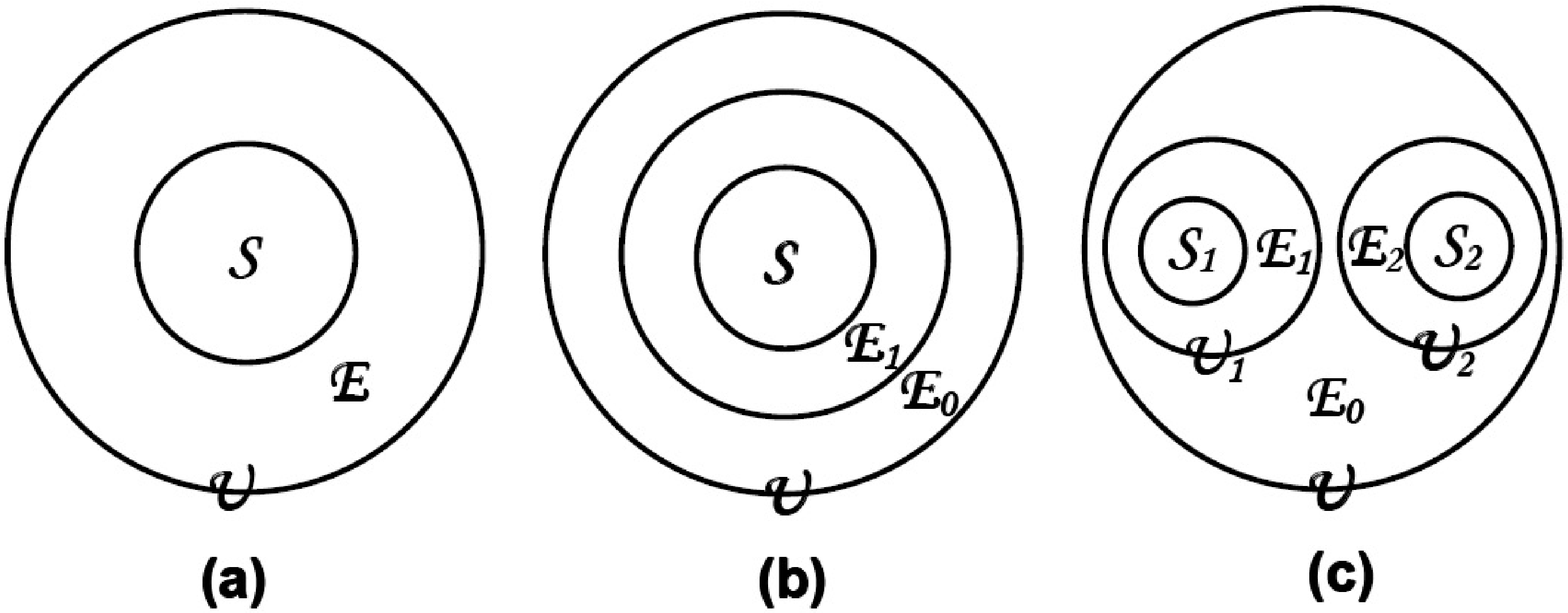}
\caption{The openness problem in quantum mechanics. Panel (a): the
basic $\mathcal{U}$$=$$\mathcal{S}$$+E$ model; panel (b): the
multi-environment model; panel (c): the multi-world model.}
\label{fig.1}
\end{figure}

Further, relating to the tracing view (I), we discuss the concept of
{\em wave function}. According to the standard QM, the property of a
system can be wholly described by its wave function. This is
definitely right, yet, not complete. This statement relies on the
assumption that there exists the wave function of the system. Yet,
according to the concept of openness we studied above, quantum
system basically is correlated with its environment, thus, the wave
function should also include the environment even when the
environment is insignificant. Along this logic, we will eventually
get that the whole universe as a whole should be a pure state,
which, in fact, is also one assumption. Below, we study the widely
concerned model of the pure universe. Let universe $\mathcal{U}$ be
composed with system $\mathcal{S}$ and environment $E$, labeled as
$\mathcal{U}=\mathcal{S}+E$. We note that there is no need to
specialize the interaction in between. The model is depicted in Fig.
\ref{fig.1}(a). Thus,
\begin{equation}
|\psi\rangle_{\mathcal{S}} \approx
\textrm{tr}_{E}\rho_{\mathcal{U}},
\end{equation}
on the left-hand-side (LHS), $|\psi\rangle_{\mathcal{S}}$ is the
pure state of the system; on the RHS,
$\textrm{tr}_E\rho_{\mathcal{U}}$ is a decohered state. When the
role (effect) of environment $E$ is trivial, we can let them equal
in mathematics, also in physics we assume they are the same. This
kind of approximation should be nontrivial for the understanding of
QM. So, for the method of wave function, we should make clear the
conjecture (or assumption) as follows:

\begin{conjecture}
In the pure universe, there exists system which has one wave
function.
\label{con:1}
\end{conjecture}

This conjecture addresses that the concept of wave function itself
indicates the concept of openness in QM. Thus, pure quantum
mechanically, when we study some quantum theory, the starting point
of the theory should be the open system, which is just opposite to
the classical mechanics (CM). In CM, often the individual behavior
of a certain object is described, instead of the correlation with
the rest of the universe. Conjecture \ref{con:1} is one of the main
differences between QM and CM. We will address later in the last
section the differences between QM and CM in detail. In addition, we
note that the similar problem has also been studied mathematically
from a statistical view, e.g., in Ref. \cite{Popescu}.

\section{Quantum Structure}
\label{sec:2}

In our study we view entanglement and information as different
quantities, as the quantity ``discord'' indicates that there could
exist quantum information without entanglement
\cite{Ollivier,Henderson}. Entanglement describes the quantum
correlation of a system, and by information, we mean Von Neumann
entropy. Here, we do not intend to make mathematical study. For the
indication of entanglement and information to QF, there are many
progresses. For instance, Mermin stated that QM describes the
correlation without ``correlata'' \cite{mermin}, which just
demonstrated the openness of quantum theory. Bub studied QM from a
broader view, i.e., comparing to the theory of Relativity, and he
claimed that QM is the ``principle'' theory, the CBH theorem tried
to serve as the principle of QM, and QM is about quantum information
instead of wave or particle \cite{Bub}, which arouses great
interests \cite{Brassard}. Gisin viewed nature as nonlocal
fundamentally, and nonlocality does not exist in space-time
\cite{Gisin}, this observation indicates that there may exist
another kind of fundamental symmetry behind the standard QM.
Relating to {\em decoherence}, Zurek systematically studied the
relation between quantum and classical. He introduced the symmetry
{\em envariance}, and viewed it as the fundamental symmetry
\cite{Zurek}. With all these explorations, yet, there are still
primary problems remain, e.g., the physical meaning of entanglement
is not clear, one expression of this confusion is that there are too
many quantities to characterize entanglement at present. Below, we
present a new kind of picture to understand entanglement and
information, i.e., we introduce a new kind of space based on
entanglement and information via the $\mathcal{U}=\mathcal{S}+E$
bi-party model above.

\subsection{Mutual Information} \label{sec:2a}

We first study the property of information via the well known mutual
information. It is direct to introduce another system or
environment, or to divide $E$ ($\mathcal{S}$) into different parts.
In Fig. \ref{fig.1}(b) and (c), we show the two basic models. Panel
(b) shows the multi-environment model (multi-$E$), and panel (c)
shows the multi-world model (multi-$W$). Here, we should demonstrate
that one world (or universe) should contain at least one system and
one environment, i.e., one system coexists with at least one
environment, this is the result of openness. In panel (c), there are
two systems $\mathcal{S}_1$ and $\mathcal{S}_2$, then in the whole
universe there are multi-world, three kinds of environment, $E_0$,
$E_1$, and $E_2$, coexist.

For the multi-$E$ model, the wave function of the system
$\mathcal{S}$ is
\begin{equation}
|\psi\rangle_{\mathcal{S}} \approx
\textrm{tr}_{E_1}\textrm{tr}_{E_0}\rho_{\mathcal{U}}. \label{eq:3}
\end{equation}

For the multi-$W$ model, the wave function of the system
$\mathcal{S}_1$ ($\mathcal{S}_2$) is
\begin{equation}
|\psi\rangle_{\mathcal{S}_{1(2)}} \approx
\textrm{tr}_{E_{1(2)}}\textrm{tr}_{\mathcal{S}_{2(1)}+E_{2(1)}}\textrm{tr}_{E_0}\rho_{\mathcal{U}}.
\label{eq:4}
\end{equation}
We should note again, the LHS and RHS of both equation (\ref{eq:3})
and (\ref{eq:4}) are made equal in both mathematics and physics.

To be more precise, we analyze the mutual information, which is
viewed as the total correlation, for the different models. The
mutual information $\mathcal{I}$ is widely involved in the research
of QIQC, such as the discord \cite{Ollivier,Henderson}, squashed
entanglement \cite{Tucci,Christandl,Oppenheim}, conditional
entanglement of mutual information \cite{Yang} with the operational
meaning of partial state merging \cite{Horodecki}, etc. In Fig.
\ref{fig.1}(a), the basic $\mathcal{U}=\mathcal{S}+E$ model is quite
simple, as $\rho_{\mathcal{U}}$ is pure, there always exists the
bi-party Schmidt decomposition of the pure state of the universe
$|\Psi\rangle_{\mathcal{U}}=\sum_i\lambda_i|\mathcal{S}_i\rangle|E_i\rangle$,
where $|\mathcal{S}_i\rangle$ and $|E_i\rangle$ are local basis,
that is, the entanglement can always be realized by the rotation of
the basis. The mutual information is
\begin{equation}
(a): \mathcal{I}=S_{\mathcal{S}}+S_E-S_{\mathcal{S}E}=2S_E.
\label{eq:5}
\end{equation}
And the classical information is $S_E$, then the quantum information
is just $S_E$, the von Neumann entropy, which is the well known
result.

For the multi-$E$ model in Fig. \ref{fig.1}(b), the density matrix
for the party $\mathcal{S}+E_1$ is
$\rho_{\mathcal{S}E_1}=\textrm{tr}_{E_0}\rho_{\mathcal{U}}$, and the
mutual information is
\begin{equation}
(b):
\mathcal{I}=S_{\mathcal{S}}+S_{E_1}-S_{\mathcal{S}E_1}.\label{eq:6}
\end{equation}
When $E_0=\O$ or $E_0=E_1$, the multi-$E$ model reduces to the model
(a).

For the multi-$W$ model, here we aim to quantify the mutual
information between systems $\mathcal{S}_1$ and $\mathcal{S}_2$,
with environments $E_1$ and $E_2$, respectively. There can be mutual
information among any two of the four parties. From the Venn
diagram, which we do not show here, it is direct to get the mutual
information between the two systems
\begin{equation}
(c):
\mathcal{I}=\mathcal{I}(\mathcal{S}_1E_1:\mathcal{S}_2E_2)-\mathcal{I}(E_1:E_2)-\mathcal{I}(E_1:\mathcal{S}_2|E_2)-\mathcal{I}(E_2:\mathcal{S}_1|E_1),\label{eq:7}
\end{equation}
where, e.g., $I(E_1:\mathcal{S}_2|E_2)$ is the conditional mutual
information. This expression is general, and it can be reduced to
special forms under the particular cases as below:

(1). When $E_1=E_2=E_0$, the multi-$W$ model reduces to the analogy
of the multi-$E$ model in Fig. \ref{fig.1}(b).

(2). When $E_1=E_2=E\neq E_0$, (or $E_1$($E_2$)$=\O$), the mutual
information reduces to
\begin{equation}
\mathcal{I}=\mathcal{I}(\mathcal{S}_1:\mathcal{S}_2|E),\label{eq:8}
\end{equation}
which is the same with the squashed entanglement physically
\cite{Tucci,Christandl,Oppenheim}.

(3). When $E_0=\O$, that is, the state
$\rho_{\mathcal{S}_1E_1\mathcal{S}_2E_2}$ is pure, then the
three-party state $\rho_{\mathcal{S}_1E_1\mathcal{S}_2}$
($\rho_{\mathcal{S}_1\mathcal{S}_2E_2}$) is mixed, and the problem
becomes the same with case (2).

(4). When $I(E_1:\mathcal{S}_2|E_2)$ and $I(E_2:\mathcal{S}_1|E_1)$
are {\em zero}, which can be easily depicted via the Venn diagram,
the mutual information in equation (\ref{eq:7}) reduces to
\begin{equation}
\mathcal{I}=\mathcal{I}(\mathcal{S}_1E_1:\mathcal{S}_2E_2)-\mathcal{I}(E_1:E_2),\label{eq:9}
\end{equation}
which is physically the same with the conditional entanglement of
mutual information defined in Ref. \cite{Yang}. This indicates that
our definition is more general.

From the above study, we can see that we can use the mutual
information to characterize the information within the different
models. And from the concept of openness, the mutual information for
the multi-$W$ model in equation (\ref{eq:7}) is the most general
one, i.e., it demonstrates that if we intend to extract the mutual
information between two systems, we need to consider the
corresponding environments of the two systems. In reality, the three
environments $E_0$, $E_1$, and $E_2$ can be the same, which can
simplify the complexity of the correlation.

\subsection{Tangnet}\label{sec:2b}

\begin{figure}
\includegraphics[scale=0.37]{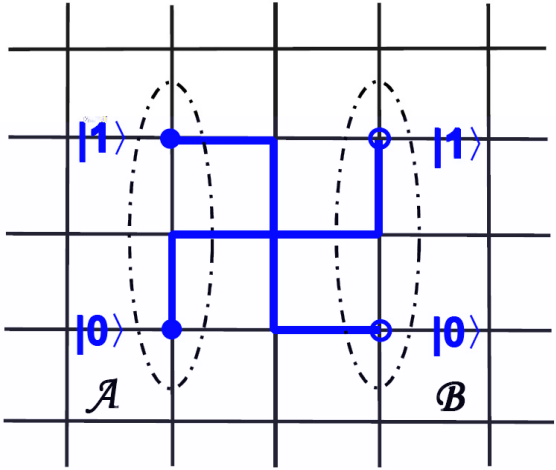}
\caption{ The tangnet space $\mathbb{T}^2$ (black lattice) and the
quantum structure of qubit (blue dots, circles, and lines). The two
parties are $A$ and $B$ (dashed-dot elliptical circles), the two
basis are $|1\rangle$ and $|0\rangle$.} \label{fig.2}
\end{figure}

We now turn to another aspect of QM. According to the standard QM,
the state vector exists in the Hilbert space. The mathematical
element in QM is operator or algebra, instead of number, that is, QM
describes the logical structure of the state. In the Heisenberg
picture, the commutation relation of the operator and the related
group can manifest the algebraic structure better than the
Schr\"odinger picture. Referring to openness, we can say that QM
describes the correlation and information of a certain dynamics.
Along the logic of the study in the above subsection, the whole
universe can be eventually depicted as a kind of ``lattice'', with
entanglement and information within. We can name this kind of space
as {\em infornet} (information-net) or {\em tangnet} (tangle-net),
shown as the lattice in Fig. \ref{fig.2}. Mathematically, the
tangnet is the topological two-dimensional complex lattice space
$\mathbb{T}^2$. We note that it is easy to put the multi-$E$ and
multi-$W$ models on the tangnet.

Tangnet is different with other spaces we are familiar with. For
instance, the configuration space or Cartesian space $\mathbb{R}^3$
in classical physics describes the possible places of the object,
which is static without time. The Minkowski space $\mathbb{M}^4$ is
the generalization of $\mathbb{R}^3$ as the result of Relativity (we
do not study the relation between QM and Relativity in this paper).
$\mathbb{M}^4$ can enclose the motion of {\em field}, which is
exotic for $\mathbb{R}^3$, by putting time and space on the equal
footing. The phase space is primarily different with $\mathbb{R}^3$
and $\mathbb{M}^4$. Phase space combines the object (its place {\bf
r}) and its movement (the momentum {\bf p}$=m\frac{d {\bf r}}{d t}$
and time) together, thus it can describe the motion more
systematically. The Hilbert space $\mathbb{H}$ is the space of state
vector, it defines the operation of operator and vector. The tangnet
$\mathbb{T}^2$ is not the differential manifold, which is the
central feature of this space. It should be interesting that this
feature can be viewed as the origin of ``quantisation'', with
manifold as the classical limit. Tangnet $\mathbb{T}^2$ is not the
same with the Hilbert space $\mathbb{H}$. In $\mathbb{T}^2$, the
state is described as the ``node'' on the lattice instead of a kind
of vector, and the lines between nodes cannot be described in
$\mathbb{H}$.

From the new concept of tangnet, it is not enough to say that
quantum system is open or closed, instead, the basic object becomes
{\em quantum structure}, which is formed by the states of the
system. QM describes the quantum structure in tangnet. Information
is stored and shared in the unique and holistic quantum structure,
which indicates that the primary feature of information is {\em
sharing}.

Below, we introduce some symbols and rules of the quantum
structure:\\

($1$). Label the eignstate as dots ``$\bullet$'' or
circles ``$\circ$'' with each party the same symbol;\\

($2$). Label the entanglement as string ``-'', the length of string
relates to the coefficients in the entangled
state;\\

($3$). There is no restriction on the spatial orientation of
$|0\rangle$, $|1\rangle$, $\cdot \cdot \cdot$ $|n\rangle$ of
each party;\\

($4$). The states of different parties correlate with each other
one-to-one;\\

($5$). The phase among the branches is defined via the relative
spatial orientation.\\

We should note that for the concept of quantum structure, we
introduce ``string'' which may reflect more information than just
``state''. Next we discuss the basic quantum structure in QM. Fig.
\ref{fig.2} shows the quantum structure of the {\em qubit}, the
state is set as
\begin{equation}
|\psi\rangle_{AB}=\frac{1}{\sqrt{2}}(|0_A 1_B\rangle-i|1_A
0_B\rangle), \label{eq:5}
\end{equation}
where the two parties are $A$ and $B$, there are two branches $|0_A
1_B\rangle$ and $|1_A 0_B\rangle$, which have the same length, thus,
equally weighted as $\frac{1}{\sqrt{2}}$. It takes the branch $|1_A
0_B\rangle$ $90^\circ$ anticlockwise to rotate to the same
orientation with the other branch $|0_A 1_B\rangle$, thus, the
relative phase between them is $-i$.

We note that the study here can also be generalized to the general
basis
\begin{eqnarray}
|\psi\rangle_{AB}^A &=& \cos \alpha |0_A 1_B\rangle+\sin \alpha
e^{i\theta}|1_A 0_B\rangle),
\\ \nonumber |\psi\rangle_{AB}^S &=& \cos \alpha |0_A
0_B\rangle+\sin \alpha e^{i\theta}|1_A 1_B\rangle),
\end{eqnarray}
where $|\psi\rangle_{AB}^A$ is asymmetric, and $|\psi\rangle_{AB}^S$
is symmetric. Here, e.g., symmetric means $|0(1)\rangle$ relates to
$|0(1)\rangle$. When $\alpha=45^\circ$, and $\theta=0^\circ,
180^\circ$, the Bell basis is realized. When $\alpha=45^\circ$,
$\theta=270^\circ$, $|\psi\rangle_{AB}^A$ reduces to the state in
equation (\ref{eq:5}). Different states of qubit have different
quantum structure. It is easy to verify that there are totally eight
kinds of structures relating to qubit (we do not present them here).
In addition, for more complicated entangled state, there are more
parties and more branches.

In QIQC, qubit is viewed as the element of quantum information,
``ebit'', which equals to one qubit plus one bit. Here, we
demonstrate that qubit is the basic quantum structure, not just from
the information-theoretic view.

Next, we study the property of the structure of qubit. Actually,
this problem has been studied quite widely. For instance, Zurek
introduced ``enviarance'' \cite{Zurek}, from which he studied the
Born's rule, we do not analyze this subject in detail. This symmetry
states that the local unitary transformation $U_A$ and $U_B$ cannot
change the global property of the entangled state
$|\psi\rangle_{AB}$, which is
\begin{equation}
U_A U_B |\psi\rangle_{AB}=|\psi\rangle_{AB}.
\end{equation}
For another line of research, this problem is often mentioned as
exchange/permutation symmetry, which is studied mainly
mathematically, e.g., in Ref. \cite{toth}. Also, we should pay
attention that the permutation symmetry has already been studied
well in quantum field theory (QFT) decades ago yet without
entanglement. Here again, we focus on the physical implication of
this quantum structure. There exist two classes of basic
operations:\\

{\em a. Local base rotation (e.g., flipping)}.

For instance, for $A$, if $|0\rangle \rightarrow
|0\rangle+i|1\rangle$, $|1\rangle\rightarrow |0\rangle-i|1\rangle$,
then for $B$, $|0\rangle \rightarrow |0\rangle+i|1\rangle$,
$|1\rangle\rightarrow |0\rangle-i|1\rangle$.\\

{\em b. Permutation (or mirror/specular reflection)}.

This transformation, i.e., exchange the states of $A$ and
$B$ correspondingly, causes nothing or a global phase change.\\

We can draw the conclusion that the quantum structure is invariant
under the unitary transformation.

We assume the validity of this symmetry without making any further
proof mathematically. Instead, we discuss what this symmetry may
means to the foundation of QM. From the concept of openness, every
system should has the inside and outside also the surface. For
example, for the party $A$ of qubit, there are two states in it, and
$B$ is the outside. With the permutation, the qubit remains. We note
that this property relates to the identity principle, here the party
$A$ and $B$ are identical. Relating to the entangled structure, this
symmetry is a kind of ``out-in'' symmetry, that is, there is
actually no distinction between outside and inside. $A$ and $B$
connect with each other in such a coherent way that they become one
unique entity without boundary, i.e., the quantum structure, and the
information is shared commonly. This out-in symmetry is the
universal and elementary symmetry in the tangnet space which has
never been demonstrated before.  Below, we set the theorem of this
symmetry.

\begin{theorem}
In quantum mechanics, there exists the out-in symmetry in the
tangnet, under which the entangled quantum structure is invariant.
\label{th:1}
\end{theorem}

In addition to this theorem, we need to quantify the quantum
structure via entropy, entanglement, etc, which we do not study
here. This theorem relates to the Conjecture \ref{con:1} above.
There seems a kind of confliction with the superposition for the
one-body system. However, relating to the Conjecture \ref{con:1}
above, there is no one-body problem in QM, the simplest case should
be two-party system, that is, the out-in symmetry demonstrated by
Theorem \ref{th:1} acts always. Thus physically, there is no
confliction between the one-body superposition and multi-party
entanglement. We give two simple examples to illustrate this point.

The first one is the double-slit interference of electron. As we
know, the state of electron is the superposed state of the two
slits, labeled as $|r\rangle$ and $|l\rangle$. As the existence of
measurement, we need to include the apparatus. When detecting at the
slit $|r\rangle$ ($|l\rangle$), the state of the apparatus is
$|R\rangle$ ($|L\rangle$), the global state of electron and
apparatus is the entangled state. By tracing out the apparatus (when
$|R\rangle$ and $|L\rangle$ are orthogonal), we get the statistical
results. Also, we can apply the ``weak measurement'' to get both the
wave and particle properties of the electron. The interruption of
different measurements can get different information of the
entangled state, and the entangled quantum structure exists always
and the out-in symmetry acts always, too.

The second example is the Rabi oscillation of the two-level atom in
cavity. When the atom emits one photon, the atom evolves from the
excited state $|e\rangle$ to the ground state $|g\rangle$, and the
vacuum from the $|n-1\rangle$ state to the $|n\rangle$ state in the
Fock space. The global state of the atom and vacuum is entangled. We
can view the atom as in the superposed state, and the out-in
symmetry still acts.

From the above analysis, we can know that the generalization from
superposition to entanglement is nontrivial, most importantly, it
brings out the new out-in symmetry underlying QM demonstrated by
Theorem \ref{th:1}.

\section{Discussion}
\label{sec:3}

In conclusion, in this work, we briefly discussed the basic concepts
in QM due to the development of entanglement and information. We
stated that QM indicates another kind of space, tangnet (or
infornet), where exists the universal out-in symmetry and quantum
structure, e.g., the most basic one, qubit. We add that further work
should be carried out on the mathematical properties of tangnet. We
also constructed the general form of the mutual information between
two systems, i.e., equation (\ref{eq:7}) in section \ref{sec:2a}.

For the theorem \ref{th:1} we conjectured, a mathematical study is
needed, particularly, the unique definition of entanglement is
necessary. Since there are too many quantities at present, such as,
concurrence, robustness etc, we need to compare them in detail. Here
we addressed that entanglement is not information physically;
instead, it forms the element of the quantum structure.

We need to discuss a little about the differences between QM and CM.
At present, there is no standard answer to this problem. According
to the orthodox interpretation, QM and CM are connected by the
``corresponding principle'', now we often relate to decoherence and
measurement. From our study, we can infer that QM is a special kind
of description of motion different with CM, they describe motion in
different ways without referring to special scales. For QM, we
showed that it describes the information and entanglement of the
motion of a certain system. For a systematic study of various
descriptions of motion, we will present in the future.

Another point is about the identity principle and the quantum
statistics. According to the standard QM, {\em spin} is viewed as
the pure quantum quantity without classical analogy, also there
exists spin only for micro-particles. The phenomenon of
superposition is also believed forbidden on the macroscopic scale.
However, we have known that the superposition can act on the
macroscopic scale, e.g., the Schr\"odinger cat. Following the method
in this work, we may have new view of spin. There is no reason to
restrict spin in the micro-world and it is possible that there
exists spin on the macroscopic scale, one of the possibilities comes
from that we should find more physical meaning of spin different
with the traditional one.

Last, we relate to the fundamental Holographic principle
\cite{Bousso}, which deals with entropy of black hole. This
principle states that the information in a region bounded by a
casual horizon is finite in bits and proportional to the area of the
horizon. Here, in the context of the quantum structure and out-in
symmetry, there is no definite boundary, or, there can be boundary
everywhere. This physical picture should also be quite interesting.

\end{document}